\documentclass[%
 aip,
 amsmath,amssymb,
 reprint,%
]{revtex4-2}
\usepackage[normalem]{ulem}
\usepackage{graphicx}
\usepackage{dcolumn}
\usepackage{bm}
\usepackage{xcolor}
\usepackage[utf8]{inputenc}
\usepackage[T1]{fontenc}
\usepackage{mathptmx}
\usepackage{etoolbox}
\usepackage{comment}

\makeatletter
\def\@email#1#2{%
 \endgroup
 \patchcmd{\titleblock@produce}
  {\frontmatter@RRAPformat}
  {\frontmatter@RRAPformat{\produce@RRAP{*#1\href{mailto:#2}{#2}}}\frontmatter@RRAPformat}
  {}{}
}%
\makeatother
\begin{document}

\preprint{AIP/123-QED}

\title[]{
A Brillouin light scattering study of the  {spin-wave} magnetic field dependence in a magnetic hybrid system made of an artificial spin-ice structure and a film underlayer}

\author{F. Montoncello}%
\affiliation{ 
Dipartimento di Fisica e Scienze della Terra, Università di Ferrara, Ferrara, Italy
}%

\author{M. T. Kaffash}
\affiliation{Department of Physics and Astronomy, University of Delaware, Newark, Delaware 19716, USA
}%

\author{H. Carfagno}
\author{M. F. Doty}
\affiliation{Department of Materials Science and Engineering, University of Delaware, Newark, Delaware 19716, USA
}%

\author{G. Gubbiotti}%
\affiliation{Istituto Officina dei Materiali del Consiglio Nazionale delle Ricerche (IOM-CNR), c/o Dipartimento di Fisica e Geologia, Perugia I-06123, Italy}

\author{M. B. Jungfleisch}%
 \email{mbj@udel.edu}
\affiliation{Department of Physics and Astronomy, University of Delaware, Newark, Delaware 19716, USA}

\date{\today}

\begin{abstract}
{We present a combined Brillouin light scattering and micromagnetic simulation}
{investigation of the magnetic-field dependent spin-wave spectra in a hybrid structure made of permalloy (NiFe)} artificial spin-ice (ASI) systems, composed of stadium-shaped nanoislands, deposited on the top of a unpatterned permalloy film with a nonmagnetic spacer layer. 
The thermal spin-wave spectra were recorded by Brillouin light scattering (BLS) as a function of the magnetic field applied along the symmetry direction of the ASI sample. Magneto-optic Kerr effect magnetometry was used to measure the hysteresis loops in the same orientation as the BLS measurements. The frequency and intensity of several spin-wave modes detected by BLS was measured as a function of the applied magnetic field. Micromagnetic simulations enabled us to identify the modes in terms of their frequency and spatial symmetry and to extract information about  {the existence and strength of the dynamic coupling, relevant only to a few modes of a given hybrid system}.  
Using this approach, 
 {we suggest a way to understand if  dynamic coupling between ASI and film modes is present or not}, 
with interesting implications for the development of future three-dimensional magnonic applications and devices.

\end{abstract}

\maketitle

\section{\label{sec:intro}Introduction}

Artificial spin ice (ASI), arrays of nanopatterned magnetic materials, offer rich physics owing to a variety of magnetic ordering in the lattice \cite{Zhou_Adv2016,Iacocca_PRB2016,Gliga_PRL2013,Jungfleisch_PRB_Ice_2016,Bang_2019,Bhat_PRB_2018,Bhat_PRB2016,Branford_PRB_2019_2,Branford_PRB_2019,Mamica_2018,Li_JPD2016,Lendinez_Nano_2021}. These systems are particularly attractive for realizing reconfigurable magnonic crystals and devices. The reconfiguration of the magnetic ordering can be driven by an external magnetic field stimulus, which in turn enables the manipulation of the spin-wave propagation characteristics, i.e., the peculiar profile of the spin-wave (SW) dynamics. 
The dynamics profile can be more or less uniform, confined in a restricted area of the primitive periodic cell \cite{APLMaterials_2022}, or extended over a larger area. The specific dynamics of a spin-wave mode determine the Brillouin light scattering (BLS) intensity, the mode bandwidth, and the group velocity \cite{Bhat_2020,Li_JPD2016,Li_JAP2017,Mamica_2018}. A thorough understanding of how the  magnetic field affects the dynamics enables precise control of the entire system dynamics. There has been significant progress in studies of the dynamics in two-dimensional lattices. More recently, studies have transitioned to the exploration of three-dimensional systems \cite{Gubbiotti_book} such as magnetic multilayered systems and systems in which unpatterned films interface {with} patterned nanostructures (e.g., ASI) \cite{Iacocca_underlayer,APLMaterials_2022}. These studies address the question of how the vertical coupling can influence the in-plane spin-wave propagation properties either through the static interaction (generating a non-uniform internal effective field distribution, $B_\mathrm{eff}$) or the dynamic interaction (generating hybridized states between the magnetic oscillations in distinct layers).

Here, we {use both BLS and micromagnetic simulations to} investigate how the biasing magnetic field affects the spin-wave dynamics in a permalloy hybrid structure comprising an ASI lattice patterned on a continuous permalloy film with nonmagnetic spacer layers of varying thickness. The ASI lattice is formed by elongated nanoelements (``islands'', in the following) to mimic  {the bistability of atomic spins in crystalline spin ice.} 
{The results presented here go beyond our previous work\cite{APLMaterials_2022}, where the ``dispersion curves'' of spin waves at a particular field value were studied. We, here, address the question of how the eigenmodes evolve as a function of the externally applied field for a fixed wavevector. 
In particular, we discuss the field dependence of both static and dynamic magnetization and derive the concept of
dynamic coupling (i.e., mode coupling, and not static dipolar interaction) from the field evolution of the
modes, illustrating how the same system can exhibit modes showing coupling while others do not. 
This is an emerging coupling mechanism that has implications for designing devices
for wave-based computing.}

The observed field-dependent dynamics are compared to the static magnetization configuration obtained by magneto-optic Kerr effect (MOKE) measurements.  We reveal how the ASI microstate affects the film, both in its static and dynamic properties. Furthermore, we demonstrate how the SW properties depend on the static coupling (through $B_\mathrm{eff}$) and the dynamic coupling of dipolar nature (through a hybridization of modes in the layers). The role of the coupling strength is determined by varying the separation between film and ASI layers using a nonmagnetic spacer layer. We observe rich BLS spectra in which we analyze the frequency intervals where the coupling between ASI and film is present as indicated by typical crossing/anticrossing behaviors. The experimental results  are interpreted using micromagnetic simulations confirming the formation of hybridizations between ASI and film modes.
While our results are derived from a specific ASI/film hybrid system, they have broad implications for more complex three-dimensional systems and {suggest general strategies for the investigation of such systems}.

\begin{table}
\begin{ruledtabular}
\begin{tabular}{cccc}
  Sample & Film (nm) & separation (nm) & ASI (nm)\\
\hline
  \#1 & 15 & -- & --\\
  \#2 & -- & -- & 15\\
  \#3 & 15 & 2 & 15\\
  \#4 & 15 & 10 & 15\\
\end{tabular}
\end{ruledtabular}
\caption{Overview of the geometric parameters of the investigated samples:
NiFe continuous (unpatterned) film (Sample \#1), isolated ASI made by NiFe islands (Sample \#2), hybrid film plus ASI systems with different thicknesses of the nonmagnetic spacer (Samples \#3 and \#4).
 \label{table}}
\end{table}

\begin{figure}[b]
\centering \includegraphics[width=6.5cm]{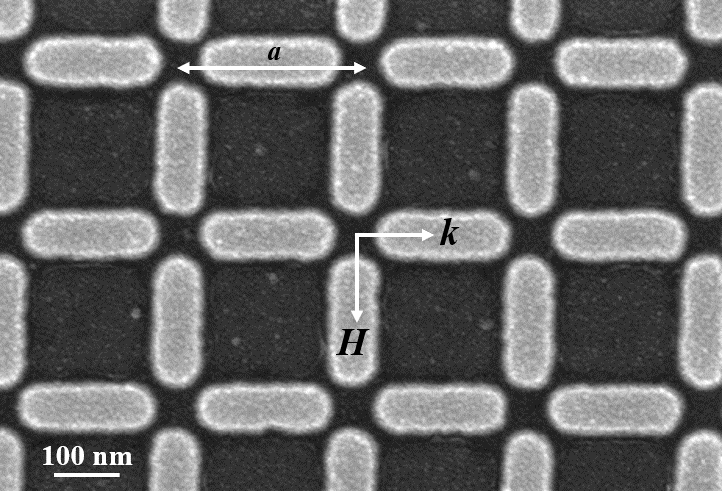}
\caption{Scanning electron microscopy (SEM) image of the artificial spin ice structure. White arrows indicate the reference system together with the direction of the  {(positive) applied magnetic field ($H$) and of the in-plane component of the light transferred wavevector ($k$). The lattice constant $a$ is approximately 355 nm.}\label{fig:SEM}} 
\end{figure}

\begin{figure}[ht]
\centering \includegraphics[width=0.7\columnwidth]{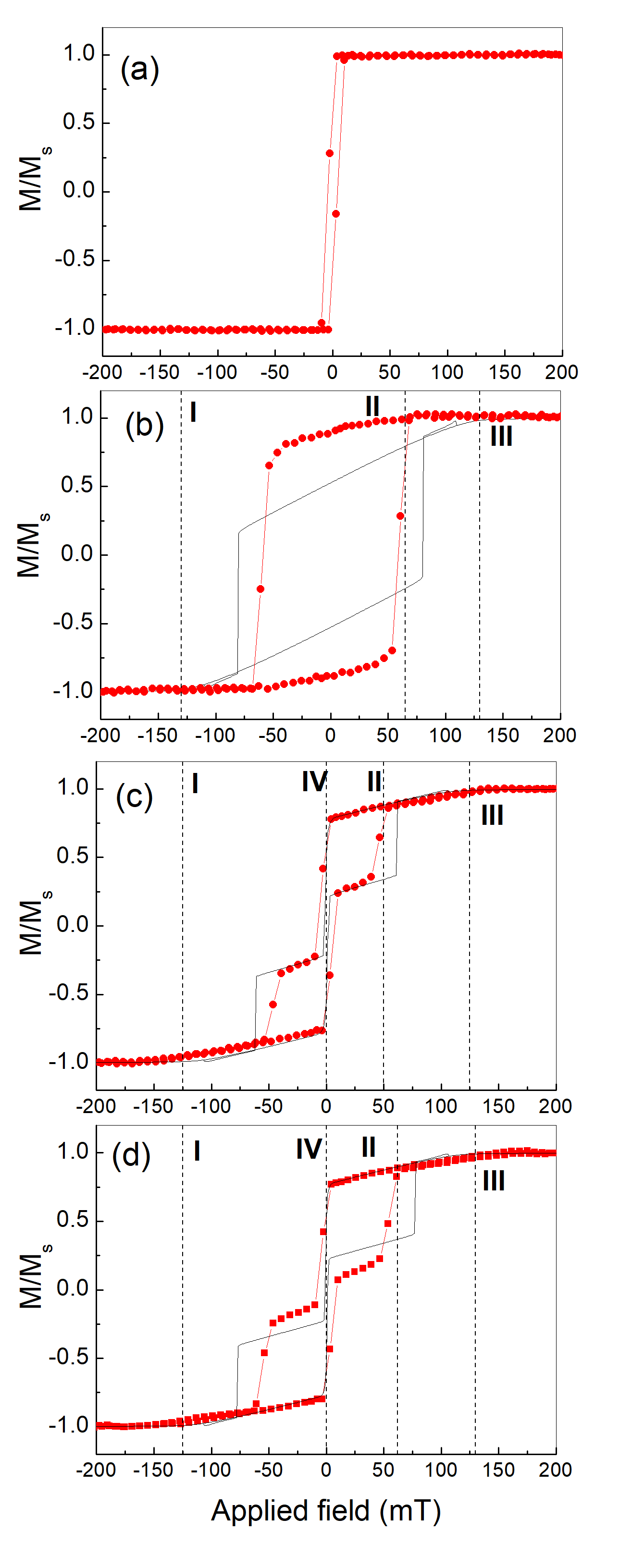}
\caption{
Measured (red symbols) and calculated (black line) hysteresis loops for (a) uncoupled film Sample~\#1, (b) uncoupled ASI Sample~\#2, (c) coupled ASI/film Sample \#3, (d) coupled ASI/film Sample~\#4. The vertical lines  {with roman numbers} mark the notable field points discussed in the text.}
\label{Fig:Kerr_loops}
\end{figure}

\section{\label{sec:details}Experimental approach and micromagnetic framework}

\subsection{Sample fabrication and characterization}

A multi-step lithography process was used for sample fabrication. First, we define the alignment marks on a thermally oxidized Si substrate using optical lithography, electron-beam evaporation of 10~nm Ti and 50~nm Au,  and lift-off. Second, we defined different underlayer/ASI structures by a combination of optical lithography, electron-beam lithography, and electron-beam evaporation. The order in which these fabrication steps are performed is varied to create the four test structures listed in Table~\ref{table}:

(a) continuous Ni$_{81}$Fe$_{19}$ film (15~nm) (Sample \#1), (b) square ASI made of 15-nm thick Ni$_{81}$Fe$_{19}$ without underlayer film (Sample \#2), (c) Ni$_{81}$Fe$_{19}$ (15~nm) with a 2~nm-thick naturally oxidized spacer layer and a square ASI made of 15-nm thick Ni$_{81}$Fe$_{19}$, and (d) Ni$_{81}$Fe$_{19}$ (15~nm)/Al$_{2}$O$_{3}$ (10~nm) with a square ASI made of 15-nm thick Ni$_{81}$Fe$_{19}$. ASI elements are composed of stadium-shaped nanoislands with lateral dimensions of $260\times90$~nm$^{2}$ arranged into a square lattice  {with lattice constant 355~nm}. The film/ASI samples \#3 and \#4 consist of the same ASI structure as Sample \#2. A representative scanning electron microscopy image is shown in Fig.~\ref{fig:SEM} (here shown Sample \#2). 
Longitudinal hysteresis loops were measured by  MOKE magnetometry at room temperature using a photoelastic modulator operating at 50 kHz and lock-in amplification.
The external magnetic field, applied parallel to the vertical islands ($y$-direction as shown in Fig.~\ref{fig:SEM}), was swept in the field range between -200 mT to +200 mT.
Thermally excited spin-wave spectra were measured in the backscattering configuration by BLS in the same magnetic field range and orientation.
For this purpose, a monochromatic laser ($\lambda$ = 532 nm) with 200 mW output power is focused onto the sample surface with a spot size of about 30 $\mu$m in diameter. 
The inelastically scattered light was analyzed in frequency using a (3+3)-pass tandem Fabry-P\'erot interferometer. We measure the frequency dependence of the applied magnetic field strength with {the angle of incidence of the light fixed} at $\theta=10^\circ$, which corresponds to an in-plane SW wavenumber $k=0.41\times 10^7$ rad/m, where $k=(4\pi$/$\lambda)\times$ sin$\theta$. This wavevector value was found to better resolve the different BLS peaks while not changing the resonance frequency compared to $k=0$. See also Ref.~[\onlinecite{APLMaterials_2022}]. 

\subsection{Micromagnetic framework and simulations}

Micromagnetic simulations are performed to reproduce both the static and dynamic properties of the sample by using the graphic processing unit (GPU) accelerated software mumax$^3$ [\onlinecite{mumax3}]. The micromagnetic system was designed as a 15-nm-thick layer onto a 15-nm-thick primitive cell of the ASI lattice and discretized in $4.0625\times4.0625\times5$~nm$^3$ micromagnetic cells. Even though the real islands are stadium-shaped, we found a better agreement using ASI elements modelled as ellipses with dimensions $64\times24$ cells, i.e., $260\times97.5$~nm$^2$. We attribute this to the edge oxidation of real islands, which determines an effective magnetic shape different from the geometric one  {(see also Supplemental Material)}. The primitive cell is a $88\times88$ square (i.e., $357.5\times357.5$~nm$^2$) in which two ellipses are included. We use periodic boundary conditions to simulate a large system. 
The magnetic parameters are: saturation magnetization $M_\mathrm{s}=800$ kA/m, exchange stiffness parameter $A=13$ pJ/m, and gyromagnetic ratio $\gamma =185\,$rad\,GHz/T. 


The external magnetic field is applied in the range -200 to +200 mT and it is tilted by an angle of $1^\circ$ to meet realistic conditions. The magnetization loops (which we compare to the MOKE measurements) are simulated from a saturated state in the $y$-direction in steps of 1~mT. At each field step, the relaxed state  {(found using a fictitious large damping factor $\alpha=0.9$ to speed up convergence)} is saved. The relaxed state is then used for the corresponding dynamical simulations  {(where $\alpha$ is set to zero)}: a uniform sinc pulse is applied perpendicular to the plane to determine the hybrid system's dynamics:

\[ 
b(t)=b\frac{\sin{2\pi f_0(t-t_0)}}{2\pi f_0(t-t_0)}
\] 
with $b=1$ mT and $f_0=25$ GHz. The 2D profile map of the out-of-plane magnetization of  {the ASI-film hybrid system} 
is recorded at a sampling time of 20 ps for a total duration of 20 ns. Then the space-resolved time Fast Fourier Transform (FFT) is applied to the magnetization maps  {of the individual layers of the hybrid ASI-film system}, identifying the contributions of either the ASI or the film layer. The distinct ASI and film layer spectra  {of the hybrid system} are then superposed with a weight chosen to agree with the intensity found in experiment.  {The reason for that is that the (single layer) FFT calculation is not normalized, so the relative height of the peaks between different calculations (for ASI and film) is not physical and an arbitrary weight is needed for the superposition.}
The SW spatial profile shown throughout the manuscript correspond to the real part of the FFT coefficients of each cell in the simulated primitive cell\cite{Ben_2014}. The BLS spectra are calculated by integrating the amplitude over all micromagnetic cells in the primitive cell ($k$=0), and then taking its square modulus (compare to theory in Ref.~[\onlinecite{Gubb_ell_2005}]).
 {The field step used for the dynamic simulations was not fixed, but varied from $\Delta\mu_0H=25$~mT where curves are smooth and almost linear to $\Delta\mu_0H= 5$~mT in transition regions and at turning points.}
 
 \begin{figure*}[hbt!]
\centering \includegraphics[width=1.5\columnwidth]{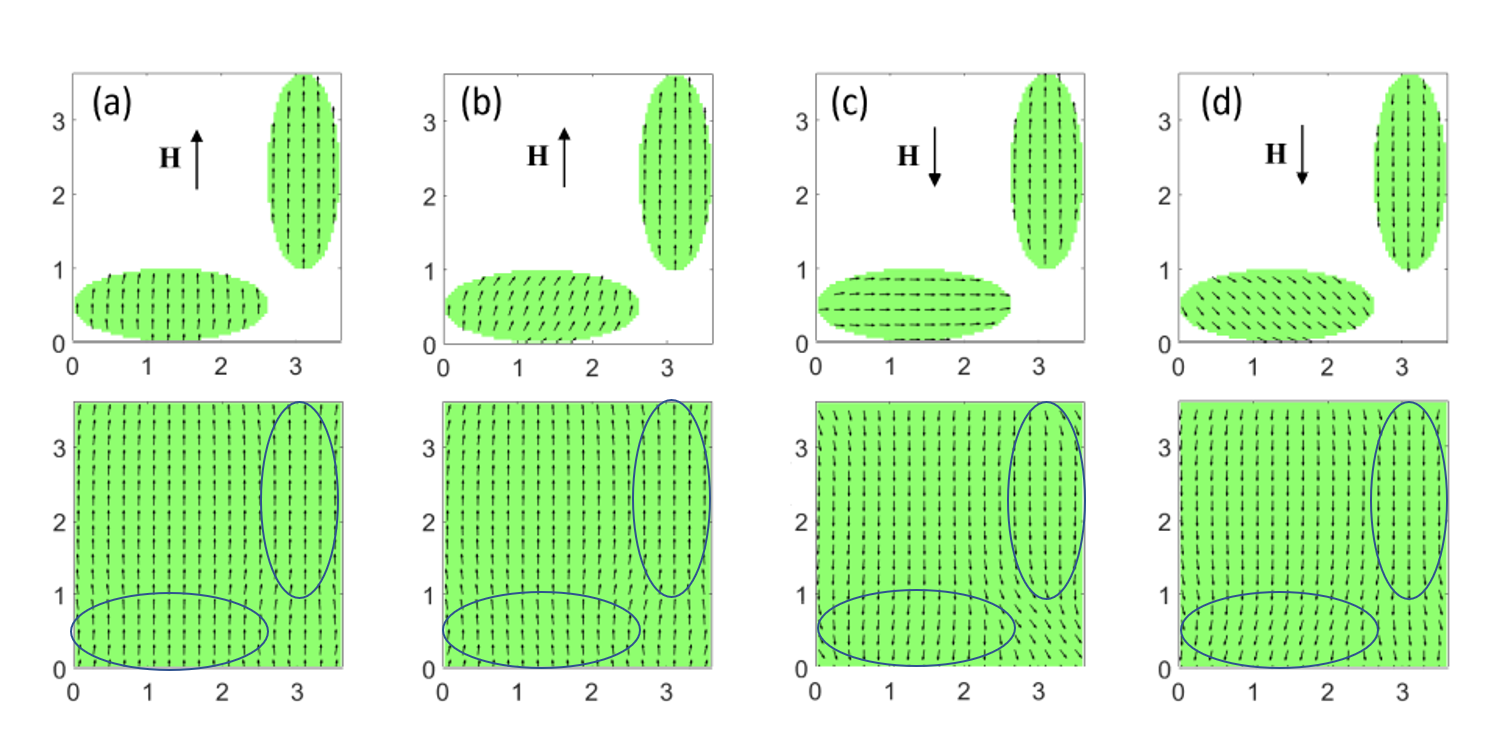}
\caption{Calculated static magnetization maps for ASI/film Sample~\#3 (top panels: ASI layer, bottom panels: film layer, {outlines in the bottom panel indicate where the ASI elements are located on the maps of the underlayer}). A similar scenario is found for Sample~\#4 (larger spacer thickness). The following field range are shown: (a) -200~mT to field line I ("onion states"); (b) maps at -100~mT, illustrative of any case after field line I (``S-states''); (c) +10~mT illustrative of region IV-II, after film reversal, i.e., field line IV; (d) +60~mT, illustrative of cases after field line II. Note that arrows are only illustrative, being each one the average of a grouping of 5 micromagnetic cells.  {The units on the axes are $\times100$ nm}. 
\label{Fig:magnetizations}}
\end{figure*}

\begin{figure*}[t!]
\centering \includegraphics[width=1.3\columnwidth]{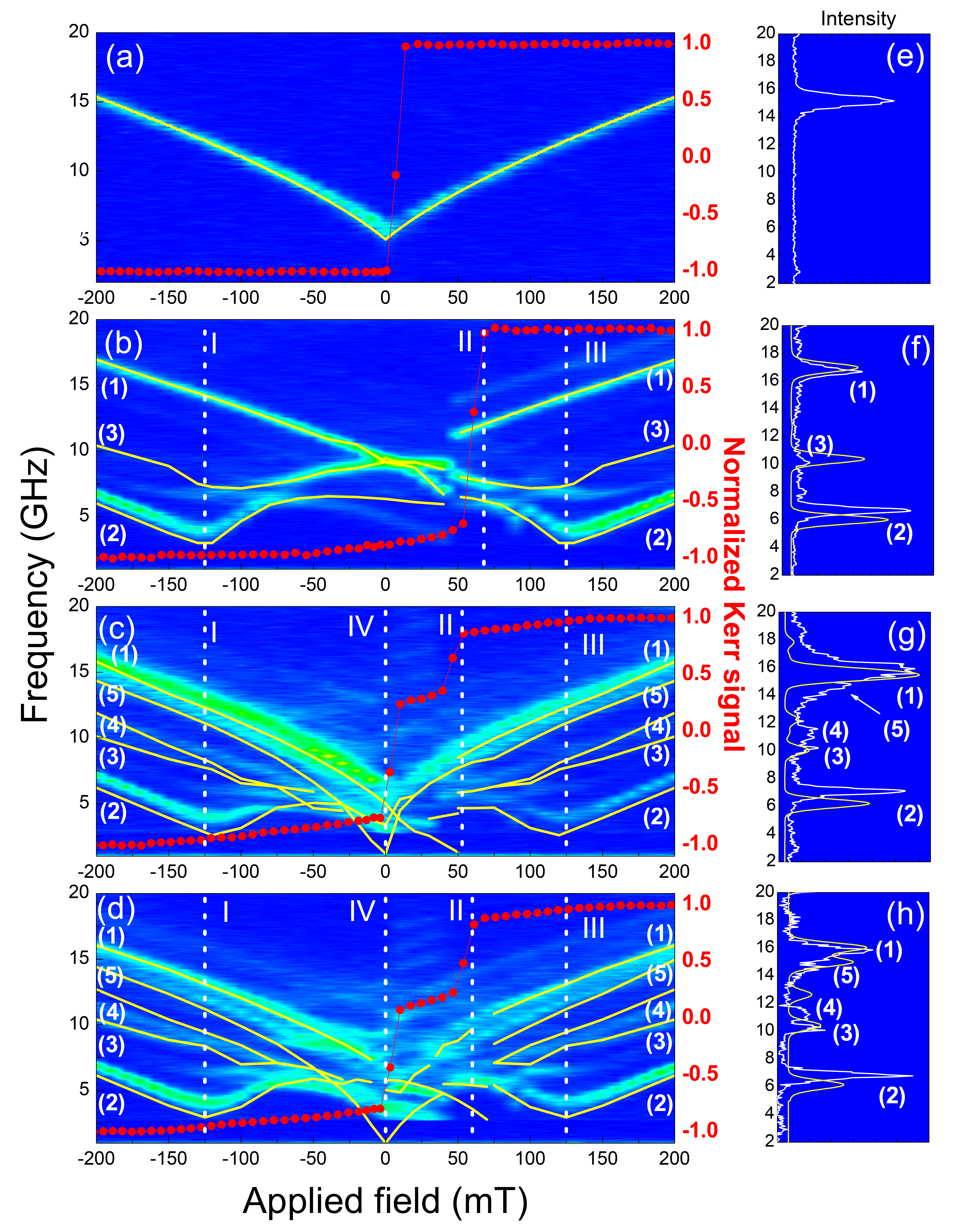}
\caption{Field dependence (increasing field from left to right) of the BLS spectra from thermally excited SWs measured at  {$k=0.41\times 10^7$~rad/m} (color scale) for (a) Sample $\#1$, (b) Sample $\#2$, (c) Sample $\#3$, (d) Sample $\#4$.   {In panel (a) the theoretical Kalinikos-Slavin curve calculated at $k=0.41\times 10^7$~rad/m is plotted with the measurements. In panel (b-d), simulations are carried out for $k=0$} (solid lines) are plotted together with numbers identifying the peaks in the spectra. The normalized MOKE loops (red symbols, axis on the right) are also shown for all the investigated samples. Vertical dashed lines mark the notable (transition) points discussed in the text.
Panels (e-h) show the corresponding Stokes-side of the BLS spectra (intensity vs frequency shift, same units of the vertical axis on the left) for an applied field of -200~mT, to which simulated spectra are superimposed (yellow curves).
\label{fig:Fieldscan}}
\end{figure*}

\section{\label{sec:results}Results}
{In Fig.~\ref{Fig:Kerr_loops} we compare the measured MOKE hysteresis loops for all investigated samples with the simulated loops and find a reasonably good qualitative agreement.}
For the unpatterned NiFe film [Sample $\#1$, Fig.~\ref{Fig:Kerr_loops}(a)], the measured MOKE loop resembles the shape of a step function, where the magnetization is almost saturated over the entire field range, except for a narrow (<~1~mT) interval around $\mu_0H=0$, where the magnetization switches to negative values.
For the isolated ASI system [Sample $\#2$, Fig.~\ref{Fig:Kerr_loops}(b)], the loop is characterized by an initial gradual decrease of the magnetization as the field magnitude is decreased, indicating a gradual rotation of the magnetization in the horizontal islands to the easy axis (in $x$-direction), followed by a sharp jump attributed to the magnetization reversal of the vertical island (identified by the vertical line II).
In the other two cases [Sample $\#3$ and Sample $\#4$], i.e., the coupled ASI/film samples, an additional third effect is observed: the magnetization reversal of the underlying film is seen when the field is swept across 0 mT (indicated by the vertical line IV). Hence, in these two cases, there is a field region, delimited by the vertical lines IV and II, within which the magnetizations in the film and the vertical island are antiparallel. We note that such a field region occurs for both positive and negative applied magnetic fields, but for clarity lines IV and II denote the boundaries of just one such region (for positive applied fields).

 {In the measurements, line II for the ASI/film systems [Fig.~\ref{Fig:Kerr_loops}(c,d)] slightly increases, from $53$~mT to $60$~mT, and the region between line IV (i.e., $B=0$) and line II is characterized by an average relative magnetization around $M/M_\mathrm{s}=0.25$. In the simulation, this region is apparently wider. This mismatch between theory and experiment is due to the unavoidable distribution of shapes across the real ASI array, which tends to smooth out discontinuities. For this reason, the actual experimental field range for region IV-II is somewhat smaller than the calculated one. The same can be said for the reversal fields of the vertical island.} 

As is apparent from Fig.~\ref{Fig:Kerr_loops}, the qualitative agreement between measurements and simulations is remarkable considering the dispersion in island shape and separations between islands in the real samples. In particular, both experiment and simulation show how the region delimited by the vertical lines IV and II increases with increasing the thickness of the non-magnetic spacer: measured (simulated) line II occurs for {Sample $\#3$ at 53 mT (60~mT); for Sample $\#4$ at 60~mT (77~mT).}

This effect can be understood by inspecting the simulation results: first of all, the film layer undergo reversal across 0~mT through a step-like rotation (as opposed to gradual rotation) occurring within a range of around 2~mT: in this range, the magnetization of the film is mismatched with respect to that of the vertical island, which hence experiences strong demagnetizing fields. After the film layer switches, the magnetization texture in the vertical island is more distorted if ASI and film layers are closer to each other (as a consequence of the stronger dipolar interaction). Hence, the vertical islands {undergo magnetization reversal more easily when on top of a small spacer layer than they do when the spacer layer is larger}.
As a result, field line II occurs at a larger value for the sample with a thicker spacer.
For external magnetic fields above field line II, and, hence after the vertical island reversal, the magnetization asymptotically approaches saturation [Fig.~\ref{Fig:magnetizations}(a)].     
We further note that the system behaves symmetrically at positive and negative applied magnetic field values: indeed, after line II, the magnetization maps at $\pm\mu_0H$ are indistinguishable (180$^\circ$ rotation).

Furthermore, we address a subtle effect that is not visible in static measurements, but only in simulations, which is marked by the field lines I and III (Fig.~\ref{Fig:Kerr_loops}): at these field values ($\pm 125$~mT) a barely appreciable change of slope in the simulated magnetization curves occur, due to the reorientation of the magnetization in the horizontal islands (change of symmetry) from an onion state to a S-state at line I (and vice versa, at line III). The different symmetries of the magnetization maps can be seen in the illustrative panels (a) and (b) in Fig.~\ref{Fig:magnetizations}.  {Experimental observation of onion and S-states were discussed in Refs. ~\cite{Rave_Hubert_2000,JMMM_2002,S-state_PRL2009}.} Nevertheless, this barely appreciable change produces a remarkably visible effect in the dynamics, i.e., a definite minimum in the SW frequency-field curve corresponding to peak (2) \cite{Bang_2019,Montoncello_2008,Hertel_2003}, which is discussed below in more detail. 

 {Moreover, we observe how the measured MOKE loops for the NiFe film/ASI samples \#3 and \#4 [(Figs.~\ref{Fig:Kerr_loops}(c) and (d)], show very sharp transitions with switching fields that do not differ significantly from those of the isolated film and ASI samples [(Figs.~\ref{Fig:Kerr_loops}(a) and (b)]. This suggests that the film and ASI layers are weakly coupled only by magnetostatic interaction (as opposed to exchange interaction).}

In the following, we describe the frequency-field behavior investigated by BLS, which we show as false-color coded plots in Fig.~\ref{fig:Fieldscan}, and discuss it in relation to the simulation results. We recall that, in our BLS measurements, the incidence plane was aligned perpendicularly to the applied field direction (Damon-Eshbach (DE) configuration). To investigate the frequency vs. field dependence, we first saturate the sample magnetization with a large magnetic field of $-200$~mT along the $y$-direction [see the simulated map in Fig.~\ref{Fig:magnetizations}(a)] and then sweep the field to positive saturation ($+200$~mT) in steps of 5~mT following the ascending branch of the hysteresis loop \cite{Tacchi}.  The false-color intensity scale in Fig.~\ref{fig:Fieldscan} presents the anti-Stokes side of the measured BLS spectra (positive frequency shift of light). 
Note that the measured MOKE loops (red dots) are superimposed in the same panels (a-d) to link the changes of the magnetization configuration to the field-dependent frequency behavior of the SW modes. In addition, we show the raw BLS spectra, together with the simulated ones, at $\mu_0 H=-200$~mT in panels (e-h). 
For the unpatterned film (Sample $\#1$), only one peak is observed in the spectrum corresponding to the DE mode. Its frequency-field curve exhibits the typical V-shape behavior centered about zero field, i.e., an almost linear dependence with symmetric values for negative and positive fields. This behavior coincides with the abrupt magnetization switching observed in the MOKE loop at zero field. The DE analytical curve is plotted as a reference, calculated for consistency with the measured one but differently from the next simulations, at $k=0.41\times 10^7$~rad/m\cite{Damon,Kalinikos_1986}.

\begin{figure}[b]
\centering \includegraphics[width=1\columnwidth]{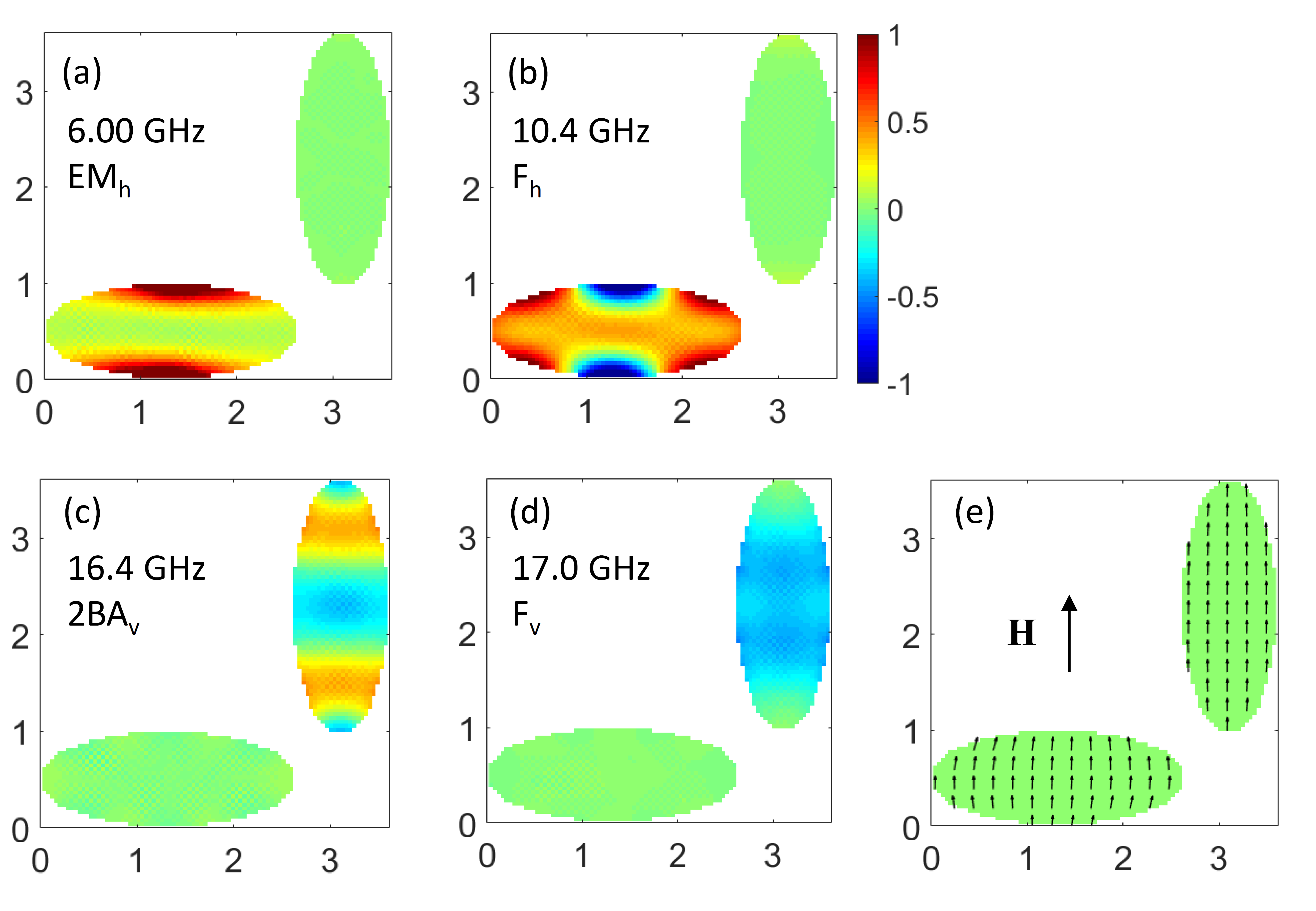}
\caption{SW phase amplitude profiles (arb. units) for the uncoupled ASI lattice, with the corresponding frequencies, calculated for $\mu_0H=$-200 mT: (a) $EM_h$; (b) $F_h$; (c) $2BA_v$; (d) $F_v$; (e) equilibrium magnetization distribution (each arrow corresponds to the average over 5 micromagnetic cells).  {The units on the axes are $\times100$ nm.} {Adapted from Ref.~[\onlinecite{APLMaterials_2022}].}\label{Fig:Mode_profilesASI}}
\end{figure}

For the ASI Sample $\#2$ shown in Fig.~\ref{fig:Fieldscan}(b), two larger intensity peaks [labeled as (1) and (2)] and a less intense one labeled as (3) are observed. Comparing these peaks to the simulations, we can associate them to the fundamental mode of the vertical island $F_v$ [peak (1)], the edge mode of the horizontal island $EM_h$ [peak (2)], and the fundamental mode of the horizontal island $F_h$ [peak (3)]. The calculated phase amplitude profiles of these modes are shown in Fig.~\ref{Fig:Mode_profilesASI}.
$F_v$ [peak (1)] belongs to the vertical island, which, due to the shape anisotropy, does not exhibit significant variation of its magnetization as the applied field magnitude is decreased to zero. This indicates an almost constant effective gyromagnetic ratio \cite{Bang_2019,Bang_PRAppl2020}, i.e., a constant slope down to zero field (Larmor-like behavior).
The frequency-field curve involving peak (2), i.e., mode $EM_h$, is characterized by the typical ``W-shape'' behavior due to the competition between the applied field and the shape anisotropy of the involved island (in this case, the horizontal one). The frequency minima at $\pm125$~mT are due to the sudden change of the slope of the magnetization curve as evidenced in Fig.~\ref{fig:Fieldscan} by field line I and III. In Ref.~[\onlinecite{Montoncello_2008}] this kind of magnetization change was studied as a second order transition, which can typically be found in any elongated nano-element (either isolated or in a lattice) initially magnetized along the short axis \cite{Montoncello_2007,Li_JAP2017,Li_JPD2016,Bang_2019,Jungfleisch_PRB_Ice_2016,Bang_JAP2019}.

\begin{figure*}[t]
\centering \includegraphics[width=1.8\columnwidth]{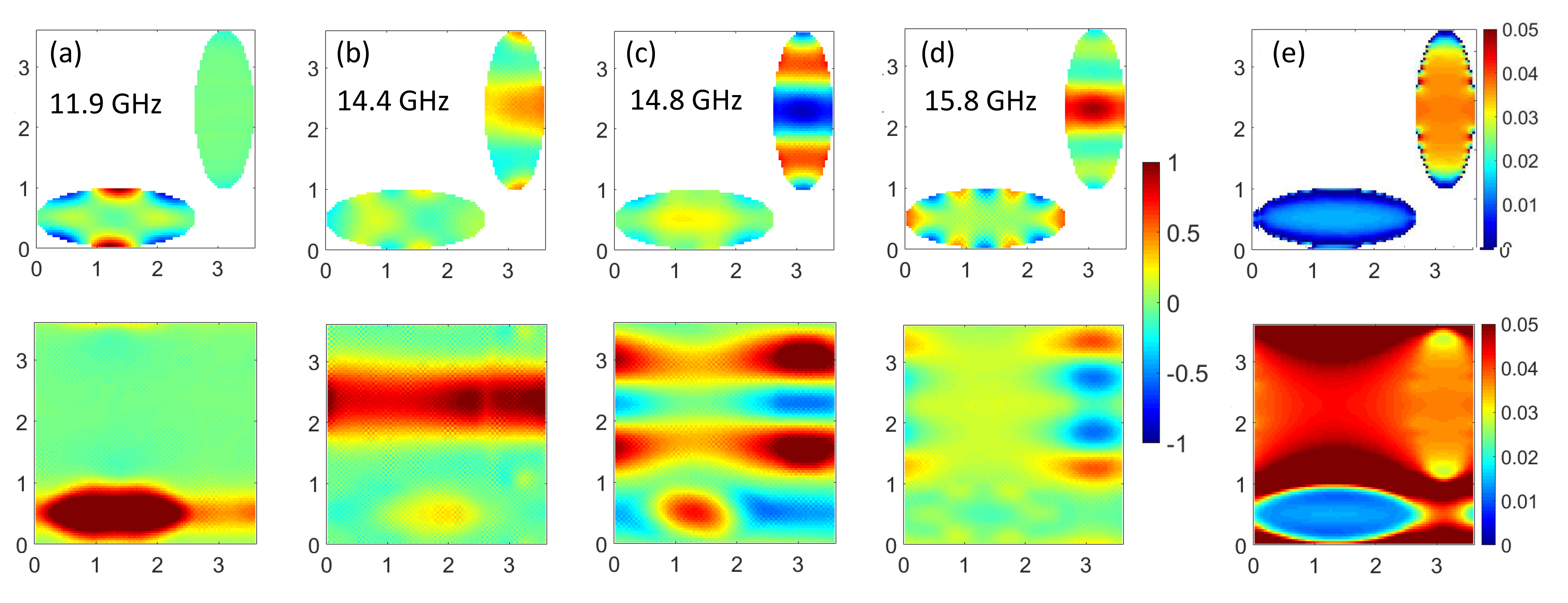}
\caption{Phase profiles (in  arbitrary units) of the coupled ASI-film system (Sample $\#3$, but indicative of Sample $\#4$ also) for $\mu_0H=$-200 mT applied along $y$: (a) mode giving the BLS peak number (4); (b) mode giving the BLS peak number (5), characterized by an ASI/film in-phase precession; (c) mode contributing to the BLS peak (1), characterized by $\lambda=a/2$ along $y$; (d) mode contributing to the BLS peak (1), characterized by an ASI/film out-of-phase precession; (e) Profile of the magnitude of $B_\mathrm{eff}$, sidebar units are T.  {The units on the axes are $\times100$ nm}. \label{Fig:CoupledModes-sp0}}
\end{figure*}

Interestingly, at around $-80$~mT, peaks (2) and (3), i.e., $EM_h$ and $F_h$ (Fig.~\ref{fig:Fieldscan}),  undergo an obvious anticrossing (``mode repulsion'', as evidenced by the formation of a `gap' in the spectrum), after which the corresponding modes exchange their character, and invert the slope of their curves (see comments in footnote~\footnote{Before -80~mT the role of the ``fundamental mode'' (i.e., no nodes, high intensity) is held by $EM_h$, while the actual $F_h$ mode has nodes (a phase change at the edges) and a lower average amplitude. After -80~mT, because the magnetization of the horizontal elements is oriented at about $45^\circ$, the $EM_h$ limits its amplitude at the edges only, while $F_h$ acquires a larger amplitude in the center.}).
The repulsion seems to be absent when the film and ASI layer couple, see Samples $\#3$ and $\#4$ [Fig.~\ref{fig:Fieldscan}(c,d)].
Close to zero field, peak (3) [which after -80~mT has gained the ``fundamental-mode'' character of previous peak (1)] intersects peak (2) at about 9.2 GHz
: this degeneracy at zero field is a clear experimental indication that the corresponding modes ($F_v$ and $F_h$) are oscillating in different regions of the sample.
  
In general, the magnetization distributions of these regions are different before the crossing point. At zero field, they  are indistinguishable, and, hence, they have the same resonance frequency.  
For the particular geometry of our ASI system, namely a periodic cell with two islands, these regions are the two differently-oriented islands in the primitive cell. 
At zero field, the two islands become indistinguishable and their magnetization is exactly aligned along their long axis.  
As a consequence, the resonances of both islands occur at the same frequency at zero field as evidenced by the observed mode intersection in the BLS curves. After $\mu_0 H=0$~mT, the curves continue their specific behavior until a frequency jump (at field line II) is observed in correspondence to the magnetization reversal of the ASI vertical islands. After this switching point, the observed behavior is symmetrical with respect to the negative field part. 
\\

\begin{figure*}[t!]
\centering \includegraphics[width=2\columnwidth]{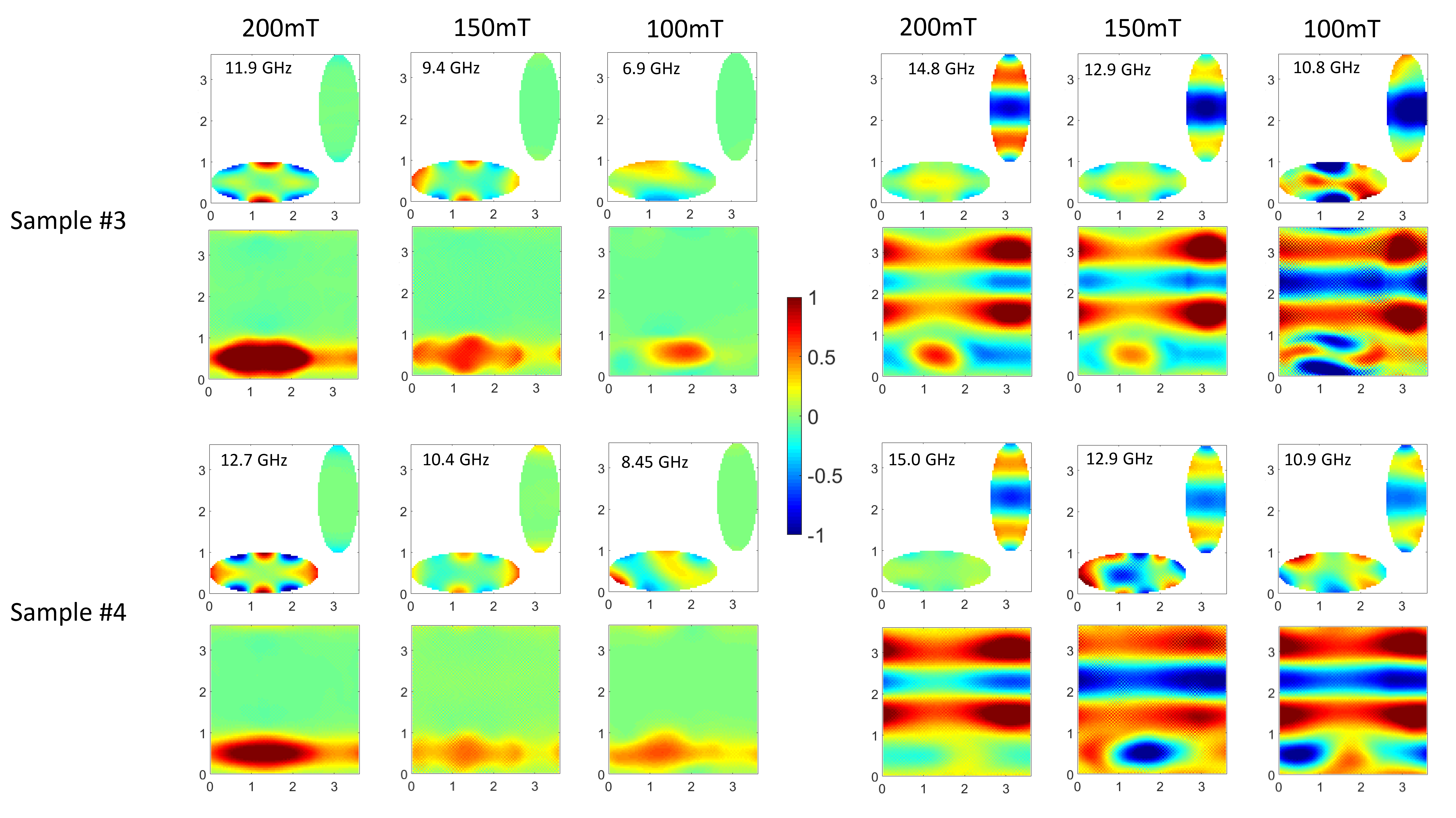}
\caption{Example of modes without (left) or with (right) dynamic coupling between layers, at different field values and spacer thickness (Sample \#3 in upper panels, and \#4 in bottom panels). For each mode, two panels are shown, corresponding to the phase amplitude profiles (in arbitrary units) in the ASI layer (top panel) and in the film layer (bottom one). In particular, on the left we plot the mode corresponding to peak (4), in which the dynamic magnetization in the ASI layer changes as field or spacer is changed (no coupling);
on the right, we plot the mode giving a large BLS peak (intermediate between peak (1) and (5)), which shows a hybridization between the mode in the ASI layer (in this case, $2BA_v$) and the one in the film layer, with a phase relationship in vertical correspondence. The association persists in changing the field value of the spacer thickness (dynamic coupling).  {The units on the axes are $\times100$ nm}.
\label{Fig:Dynamic_coupling}}
\end{figure*}

The SW dynamics in the hybrid ASI/film systems at $\mu_0H=200$~mT is reported in detail in Ref.~[\onlinecite{APLMaterials_2022}], where it was found that the ASI modes with resonance frequencies below the DE mode of the uncoupled film are almost insensitive to the presence of the film underlayer: as found in the simulations, in this frequency range the modes show in the ASI layer a phase profile almost identical to that shown in Fig.~\ref{Fig:Mode_profilesASI} for the uncoupled ASI lattice, while in the film layer the dynamic magnetization is almost negligible. 

On the other hand, in the same reference, it was found that modes close to the DE frequency can show a phase profile hybridization between ASI and film layer. This type of hybridization is 
called dynamic coupling. This effect, if present, is a robust link between a specific mode of the film layer and a specific mode of the ASI layer. The phase profiles of a few of these ASI/film modes (i.e., those with frequency close to the DE mode of the film), having a relevance in the BLS spectra, are shown in Fig.~\ref{Fig:CoupledModes-sp0}. Mode (a), occurring at 11.9~GHz for Sample $\#3$ (for $\mu_0H=\pm200$~mT), is reminiscent of the DE mode in the film. However, it now resides in the absolute minimum of the effective magnetic field well $B_\mathrm{eff}$ [within 1-3$\times100$ nm along $y$, see panel (e)] created in the film by the static dipolar interaction between the layers: this mode does not show any dynamic coupling between ASI and film layer, 
since the specific phase profile in the ASI layer changes when the external field or the spacer layer thickness is changed.
Mode (b) is as well reminiscent of the DE mode of the film, but trapped in the other (relative) minimum of $B_\mathrm{eff}$  {(within 1.5-3$\times100$ nm along $y$)}. This minimum is shallower than the previous one, and correspondingly the frequency of this mode is higher than the previous mode (a), i.e., 14.4 GHz for Sample~$\#3$ at $\pm200$~mT. As is apparent from the two-layer phase profiles, there is a vertical correspondence between the phase amplitudes (hybridization). This correspondence is robust with respect to field or spacer thickness variations (Fig.~\ref{Fig:Dynamic_coupling}): hence, this mode shows a dynamic coupling of dipolar origin between ASI and film layers.
Also in mode (c) and (d) the dynamic coupling is apparently present: in (c) the coupling results in a phase coherence between the layers, such that ASI can imprint its dynamic nodal structure in-phase in the film, while in (d) an out-of-phase relationship is imprinted.
In Fig.~\ref{Fig:Dynamic_coupling} we illustrate {the different behavior of two among the many modes} present in our system: without (panels on the left), and with (panels on the right) the dynamic coupling, or, in other terms, the possible persistence of the association between a specific ASI mode and a specific film mode in spite of biasing magnetic field or spacer layer thickness variations.
As is apparent from  Fig.~\ref{Fig:Dynamic_coupling} (panels on the left), for the mode on the left {(at 200, 150, 100~mT and, correspondingly, 11.9, 9.4, and 6.9 GHz)} [i.e., the same mode as shown in Fig.~\ref{Fig:CoupledModes-sp0}(a) giving rise to peak (4) in the experiments], the dynamic magnetization in the associated ASI layer changes when either the applied field or the spacer thickness changes. 
This means that there is no link between the excitations {of ASI and film layers.}

On the other hand, for the mode on the right {(at 200, 150, 100~mT and, correspondingly, at 14.8, 12.9, and 10.8 GHz) }[i.e., the same mode as shown in Fig.~\ref{Fig:CoupledModes-sp0}(c), contributing to peak (1) in the experiments], the association of the specific ASI mode with the specific film mode (in vertical correspondence) is persistent against any field or spacer variations. This is true within a range which preserves the general symmetry of the magnetization distribution. The range of values, of any of these parameters, preserving the coupling is itself an indirect quantification of such dynamic interaction.  {As far as this mode is concerned, we found that the coupling persists down to -75~mT, below which the profiles are hardly recognizable and the hybridization is no longer present.}

As discussed in Ref.~[\onlinecite{APLMaterials_2022}], the simulations suggest that modes (b,c,d) in Fig.~\ref{Fig:CoupledModes-sp0} with frequencies close to one another contribute to peak (1). However, we note that occasionally mode (b) can emerge in the spectra as an independent shoulder peak (5). In particular, mode (b) and (d) can be thought of as the (theoretical) extremes of the experimental broad curve of peak (1), while mode (c) has a frequency which is intermediate. This interpretation is supported by the experiment: the BLS spectra correspondingly show a broad peak {at a single field value [-200~mT in Fig.~\ref{fig:Fieldscan}(g,h)],} and the measured curve is a broad colored band in Fig.~\ref{fig:Fieldscan}(c,d) at any field value in correspondence to those modes.

In the following, we examine the detailed frequency/field curves of the hybrid ASI/film systems [i.e., Samples $\#3$ and $\#4$, Figs.~\ref{fig:Fieldscan}(c,d)]. We observe the lowest frequency curve (due to peak (2), i.e., mode $EM_h$) to behave almost the same in both coupled samples, showing the typical ``W-shape'' behavior found in the uncoupled ASI, in particular with the frequency minima at lines I-III. This observation can be easily understood by considering the following two facts:
1.) Below the DE mode frequency the ASI layer is uncoupled from the film layer: the values of $B_\mathrm{eff}$ [Fig.~\ref{Fig:CoupledModes-sp0}(e)] experienced by the low frequency modes (in the horizontal island area) is very low (dark blue) with respect to the corresponding area in the film layer (light blue). This favors decoupled magnetization dynamics, and in our specific case, only the ASI layer is excited at these frequencies.
2.) The ASI response can be understood as a superpositon of its building block's responses \cite{Bang_JAP2019}. It has been shown previously that it is possible to {derive} 
the magnetization orientation of each ASI element and the localization of the specific mode by extracting the effective gyromagnetic ratio from experimental data\cite{,Bang_2019,Bang_PRAppl2020}.

Based on these criteria, we observe that the curves associated with peaks (1) and (3-5) decrease almost monotonically, both in experiments and simulations. This frequency decrease continues until field line IV (i.e., 0~mT) when the film undergoes magnetization reversal.

\begin{figure*}[ht]
\centering \includegraphics[width=1.5\columnwidth]{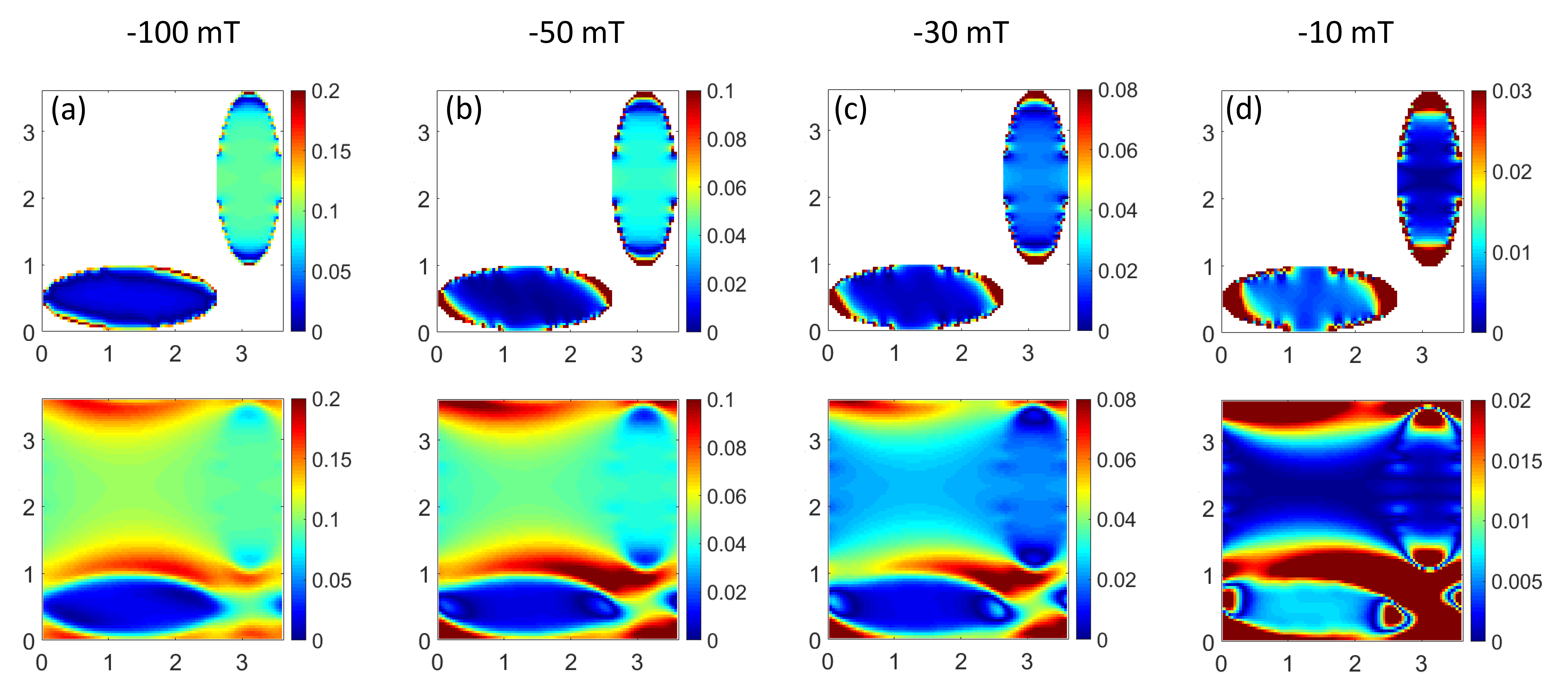}
\caption{Variation with the applied field of the space profile of $B_\mathrm{eff}$ in the film layer, (a) $-100$~mT, (b) $-50$~mT, (c) $-30$~mT, (d) $-10$~mT: the colormap refers to the $y$ component (in units of T), i.e., along $H$. Two regions form, separated by high value $B_\mathrm{eff}$ barriers (red), one within  {0-1 (units are $\times$100~nm}) is the (deepest) absolute minimum, the other within  {1.5-3.5} is a (shallower) relative minimum; however, below -30~mT their role is exchanged. {The units on the axes are $\times100$ nm.}\label{Fig:B_eff-variation}}
\end{figure*}

As mentioned above, mode (a) of Fig.~\ref{Fig:CoupledModes-sp0}, corresponding to peak (4), decreases in frequency as the field is increased from $-200$~mT. This mode gradually diminishes in intensity as the field crosses the value $-30$~mT, until the mode disappears in the spectra above $-10$~mT. This is also found in the simulations: we attribute this effect to the behavior of the two $B_\mathrm{eff}$ regions (wells). The simulations (Fig.~\ref{Fig:B_eff-variation}) reveal the deepest well as 
the one below the horizontal island (within  {0-1$\times100$ nm along $y$)} from $-200$~mT to around $-30$~mT. However, from about $-10$~mT {to higher applied field values }
the deepest well is the upper one ( {$1.5-3.5\times100$ nm }), while the former becomes shallower due to the progressive rotation of the magnetization into the long (easy) axis ($x$-direction).
Although a thorough investigation of this phenomena is outside the scope of the paper, we suggest that the reason lies in the depth of the $B_\mathrm{eff}$ well in the film in vertical correspondence to the horizontal island: at higher fields this well is not sufficiently deep to host a localized excitation because the magnetic moments are better aligned and the relative demagnetizing field smaller. 


{Interestingly, the intensity of mode (b) in Fig.~\ref{Fig:CoupledModes-sp0} [peak (5)], is rather weak at $-200$~mT [the mode appears in the spectra as a left shoulder to the main peak (1)], but, based on simulation results, this mode acquires intensity as soon as the field is increased, and above $-180$~mT it becomes the most intense one. However, this effect could hardly be detected in the BLS experiment, which always detects a broad band.}

A representative magnetization map for the range delimited by line IV and II (``IV-II region'') is shown in Fig.~\ref{Fig:magnetizations}(c) (here $+10$~mT). In this region IV-II, the magnetization in the film layer is (on average) parallel to the applied field (in $y$-direction) and as the field is further increased, it progressively becomes more aligned to the (negative) $y$-axis: the consequence is a frequency increase of the modes localized in the film layer [i.e., modes (a-d) in Fig.~\ref{Fig:CoupledModes-sp0}, and in particular the mode that gives rise to peak (5)]. 

With reference to {the simulations, we can identify four curves in this region IV-II for Samples $\#3$ and $\#4$: 
1.) the curve ascending from a minimum frequency at zero field, corresponding to the ``acoustic''-like mode (b) in Fig.~\ref{Fig:CoupledModes-sp0} is identified as peak (5) in the spectra. In the ideal case, this mode should not have any (or only a small) discontinuity in crossing 
the field line II, where the reversal of the vertical island occurs (see Ref.~[\onlinecite{Kuzma_2018}]);
2.) the curve associated with peak (1) continues to decrease even after line IV (i.e., above zero field) down to the transition point line II. This is due to the ``optical'' character of the mode (d) in Fig.~\ref{Fig:CoupledModes-sp0}. This curve experiences a strong discontinuity at the transition point near line II;
3.) and 4.) the two curves associated with $EM_h$ [peak (2)] and $F_h$ [peak (3)] in the ASI layer show negligible dynamics in the film layer. These modes are strictly localized in the horizontal islands, and their frequency {depends on the magnetization map of those islands}. Their frequency-field curves have an irregular behavior in the IV-II region. This is due to the fact that the magnetization of the horizontal island is perpendicular to the one of the film [Fig.~\ref{Fig:magnetizations}(c)]: after line IV, the magnetization first tends to align to the easy axis coming from a configuration similar to (b) of the same figure. This alignment results in a frequency increase. After that, the magnetization realigns as is shown in panel (d) of the same figure. This realignment causes a frequency decrease. Those two modes have only a small intensity due to the specific profile and are therefore difficult to resolve in the experiment.}

\section{\label{sec:summary}Conclusions}
In conclusion, we studied a ``vertical magnonic structure'' based on a hybrid ASI/film systems with varying spacer layer thickness by BLS spectroscopy. We correlate the measured dynamics with the magnetization loops measured by Kerr magnetometry and identify the main discontinuities in the loops as changes of the magnetization either in the film or the ASI layer.
Using micromagnetic simulations, we  associate each experimental BLS curve with one ore more SW modes of significant intensity. Furthermore, the simulations enabled us to understand the specific dependence on the field (slope and effective gyromagnetic ratio) with considerations about the mode profile in either layer and the underlying magnetization orientations. We reveal that the internal effective field behavior determines  the mode phase profiles (with their peculiar localizations) and the dynamic relationship between the two layers and the frequency-field evolution.
Our work identifies the role of dynamic coupling as an additional interaction  between modes of different layers, emerging in the dynamics from the geometric complexity of the structure, which preserves the associated modes in either layer in spite of any variation of applied field or spacer thickness. This coupling must be considered a hybridization between modes of different layers, persisting within certain range of values, which is an indirect quantification of this interaction.
Investigating the field-dependent dynamics resulting form the interplay between the layers and the emerging interlayer hybridization can unveil a new degree of freedom, which is particularly useful in manipulating the spin-wave intensity profile and propagation properties. The findings have direct implications for magnonic applications including spin wave interferometry, spin-wave logic gates and devices, spin-wave computing, and tuneable dynamic magnetic metamaterials.

\section{Supplemental Material}
See supplemental material for a discussion of the micromagnetic representation of the fabricated islands.

\begin{acknowledgments}
Research at the University of Delaware including sample design and fabrication supported by the U.S. Department of Energy, Office of Basic Energy Sciences, Division of Materials Sciences and Engineering under Award DE-SC0020308. This research was partially supported by NSF through ENG-1839056. G.G. acknowledges financial support from the Italian Ministry of University and Research through the PRIN-2020 project entitled “The Italian factory of micromagnetic modelling and spintronics”, cod. 2020LWPKH7. This project received funding from the European Union’s Horizon 2020 Research and Innovation Program under Grant Agreement
No. 101007417 having benefited from the IOM-CNR access provider
in the Perugia access site within the framework of the NFFA-Europe
Pilot Transnational Access Activity, Proposal No. ID151.
\end{acknowledgments}

\section{Data Availability}
The data that support the findings of this study are available from the corresponding author upon reasonable request.
\bibliography{aipsamp}

\end{document}